\documentclass[amsmath,amssymb,twocolumn,amsfonts,twocolumn]{revtex4}
\usepackage{graphicx}

\def \beq {\begin{equation}}
\def \eeq {\end{equation}}
\def \tr {\rm Tr}

\begin{document}
\title{Quantum Dynamics of Radical-Ion-Pair Reactions}
\author{I. K. Kominis}

\affiliation{Department of Physics, University of Crete, Heraklion
71103, Greece}

\begin{abstract}
Radical-ion-pair reactions were recently shown to represent a rich biophysical laboratory for the application of quantum measurement theory methods and concepts, casting doubt on the validity of the theoretical treatment of these reactions and the results thereof that has been at the core of spin chemistry for several decades now. The ensued scientific debate, although exciting, is plagued with several misconceptions. We will here provide a comprehensive treatment of the quantum dynamics of radical-ion-pair reactions, generalizing our recent work and elaborating on the analogy with the double-slit experiment having partial "which-path" information. This analogy directly leads to the general treatment of radical-ion pair reactions covering the
whole range between the two extremes, that of perfect singlet-triplet coherence and that of complete incoherence.
\end{abstract}

\maketitle
\section{Introduction}
Spin-selective radical-ion-pair reactions are a rare example in chemistry where
spin degrees of freedom and their relatively small interaction energy can have a
disproportionately large effect on the outcome of chemical reactions. The study of radical-ion-pair reactions is at the core of spin chemistry \cite{steiner}, by now a mature research field directly related to photochemistry \cite{turro} and photosynthesis \cite{blankenship}.

Radical-ion pairs are biomolecular ions created by a charge transfer from a
photo-excited D$^*$A donor-acceptor molecular dyad DA, schematically described by the reaction ${\rm DA}\rightarrow {\rm D^{*}A}\rightarrow {\rm D}^{\bullet +}{\rm A}^{\bullet -}$, where the two dots represent the two unpaired electrons. The magnetic nuclei of the donor and acceptor molecules couple to the two electrons via the hyperfine interaction, leading to singlet-triplet mixing, i.e. a coherent oscillation of the spin state of the electrons. The reaction is terminated by the reverse charge transfer, resulting to the charge recombination of the radical-ion-pair and the formation of the neutral reaction products. It is angular momentum conservation at this step that empowers the molecule's spin degrees of freedom to determine the reaction's fate: only singlet state radical-ion pairs can recombine to reform the neutral DA molecules, whereas triplet radical-ion pairs recombine to a different metastable triplet neutral product. If the lifetime of radical-ion pairs is large enough, it is seen that hyperfine and Zeeman interactions, negligible compared to thermal energy, can produce a large effect on the reaction yield.

Theoretically, the fate of radical-ion-pair reactions is accounted for by the time
evolution of $\rho$, the density matrix describing the spin state of the molecule's two electrons and magnetic nuclei. The master equation satisfied by $\rho$ currently is at the center of a scientific debate since it was shown \cite{kominis_PRE} that from the advent of spin chemistry until now, the master equation traditionally used to pursue all theoretical work was phenomenological, masking the existence of non-trivial quantum effects and leading to an incomplete understanding of radical-ion-pair reactions. A fundamental master equation was derived based on quantum measurement theory, as the radical-ion-pair recombination process was interpreted \cite{kominis_PRE} to be a continuous quantum measurement of the radical-pair's spin state.

Since almost all possible theoretical predictions that can be made on radical-ion-pair reactions are founded on the master equation for the radical pair's density matrix, it is clearly important for the scientific community to converge to a common understanding of the quantum foundations of spin selective radical-ion pair reactions. Such reactions determine the late-stage dynamics in photosynthetic reaction centers \cite{boxer1,boxer2}, and furthermore there is increasing evidence that radical-ion-pair reactions underlie the avian compass mechanism, i.e. the biochemical compass used by migratory birds to navigate through the geomagnetic field \cite{schulten1,ww1,ritz,ww2,schulten2,maeda,rodgers}. A deeper understanding of the quantum dynamics inherent in radical-ion-pair reactions will thus enhance and quite probably significantly change our understanding of these biological processes.

Recent works have, however, challenged the theoretical description developed in \cite{kominis_PRE}, sparking a debate on which is the fundamental master equation describing the spin-state evolution of reacting radical-ion pairs. In order of appearance, Jones \& Hore \cite{JH} have produced a different master equation for $\rho$, conjectured to follow
from quantum measurement theory. Shushin \cite{shushin} claimed that the Bloch-Redfield relaxation theory already accounts for the quantum dynamics of radical-ion-pair reactions and no new concept from quantum measurement theory is actually needed. Most recently, Ivanov {\it et. al.} \cite{ivanov} and Purtov \cite{purtov} put forward a derivation of the radical-ion-pair master equation, leading to the traditional equation of spin chemistry.

We will here address in detail the misconceptions contained in these recent papers and then move on to examine in depth the quantum measurement dynamics of radical-ion pair reactions, attempting to shed light on several aspects of this system and their physical interpretation. We will also address our own misconceptions, showing that our previous work \cite{kominis_PRE} describes part of the picture, and a more general theoretical description is necessary. Such a description has been developed in \cite{kom_short} where it was shown that radical-ion pair reactions are analogous to the archetypal quantum paradigm, the double slit experiment with partial "which-path" information. In the following section we outline the physical system under study, while in Section III we analyze the conceptual problems of the approaches in \cite{JH}, \cite{shushin}, \cite{ivanov} and \cite{purtov}. In Section IV we present a detailed analysis of the general theory, and among other things, we analyze single molecule experiments in order to elucidate severe inconsistencies of the traditional and the Jones-Hore theories. Finally in Section V we address the issue of energy conservation in radical-ion-pair reactions, which serves as another fundamental benchmark in the comparison of the various theoretical descriptions.
\section{Radical-ion pairs as an open quantum system}
\begin{figure}
\includegraphics[width=8.5 cm]{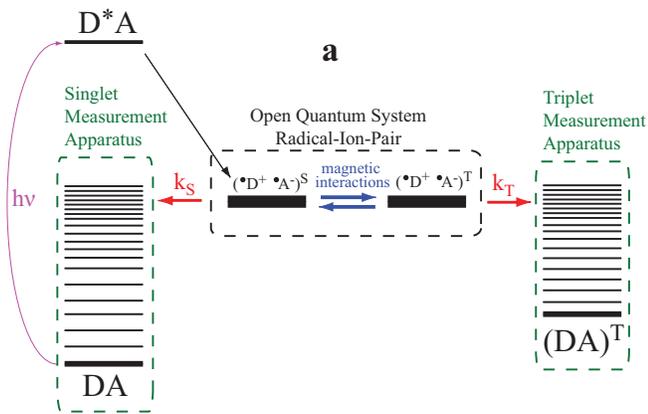}
\caption{Energy level structure and reaction dynamics of radical-ion pairs. A photon excites the singlet neutral precursor molecule DA into D$^*$A, and a charge transfer creates the radical-ion pair.
The excited vibrational levels (DA)$^*$ of the neutral DA molecule form the measurement reservoir, which has two functions: it acts as a measurement device for the spin state of the radical-ion pair, and it acts as a sink of radical-ion pairs, i.e. if a radical-ion pair recombines into the singlet channel, the electron tunnels into a reservoir state , and a fast spontaneous decay results in the ground state DA (which is the singlet product) and a photon emission. Similar for the triplet reservoir. }
 \label{fig1}
\end{figure}

We will first describe in detail the physical system under consideration. The energy levels and reaction dynamics of radical-ion pairs are depicted in Fig.\ref{fig1}. We do not consider the dynamics of creation of radical-ion pairs and we assume a given initial condition for the pair's spin state. For example, in many cases radical-ion pairs are created in the electronic singlet state and the nuclear spins are completely random, i.e. they are in the maximally mixed state. What we care about in this work is the combined dynamics of (i) the coherent singlet-triplet mixing induced by the magnetic interactions and (ii) the irreversible spin-state-dependent charge recombination that terminates the reaction.
For the study of the quantum dynamics of the recombination process, we consider a radical-ion-pair with its center of mass fixed. In other words, we completely neglect effects stemming from diffusion in solutions, collisions with other molecules, spin relaxation etc, since at this point we care to understand the fundamental quantum dynamics inherent in the radical-ion pair. Now, in all experimental studies we have a macroscopic number of radical-ion pairs. Each one of them is a single open quantum system regarding the spin degrees of freedom. It is an open quantum system because the recombination process inherent in each molecule disturbs a what would be a unitary spin evolution. Moreover, due to recombination, radical-ion pairs disappear from the ensemble and are replaced by neutral chemical products. That is, what we deal with is an ensemble of open quantum systems with varying number of particles. This non-trivial aspect of radical-ion pair reactions and the confusion stemming thereof is in part fuelling the recent debate on the fundamental master equation describing these reactions.
\section{Theoretical Description of Radical-Ion-Pair Reactions}
Since the advent of the field of spin chemistry, the spin chemistry community has described the evolution of the spin state of radical-ion pairs by the master equation (from now on to be called "traditional")
\beq
{{d\rho}\over {dt}}=-i[{\cal H},\rho]-{k_{S}\over 2}(\rho Q_{S}+Q_{S}\rho)-{k_{T}\over 2}(\rho Q_{T}+Q_{T}\rho)\label{phen}
\eeq
where the unitary evolution of the radical-pair's spins state is driven by the magnetic interactions embodied in ${\cal H}$, and the other terms attempt to represent the reaction taking place through the singlet (rate $k_S$) or the triplet (rate $k_T$) recombination channel. A few definitions are in order. The dimension of the density matrix and all operators acting on the radical-ion pair's Hilbert space is $4n$, where 4 is the spin multiplicity of the two electrons and $n=(2I_{1}+1)(2I_{2}+1)...(2I_{k}+1)$ is the nuclear spin multiplicity of the $k$ magnetic nuclei (residing in the donor and acceptor molecules) and having nuclear spins $I_1$, $I_2$,...,$I_k$. The operators $Q_{S}$ and $Q_{T}$ are the singlet and triplet projection operators, given by
\begin{align}
Q_{S}&=1/4-\mathbf{s}_{1}\cdot\mathbf{s}_{2}\\\nonumber
Q_{T}&=3/4+\mathbf{s}_{1}\cdot\mathbf{s}_{2}
\end{align}
These two projector operators add up to unity, $Q_{S}+Q_{T}=1$, i.e. the radical-ion pair is either singlet or triplet.

In 2009 Kominis \cite{kominis_PRE} argued that this master equation misses a lot of the physics of the recombination process. By properly identifying the quantum system and the environment  "watching" the system's evolution, the master equation describing the evolution of the radical-ion-pair's spin state {\it until it reacts} was derived in \cite{kominis_PRE}:
\beq
{{d\rho}_{\rm nr}\over {dt}}=-i[{\cal H},\rho]-{{k_{S}+k_{T}}\over 2}\big(\rho Q_{S}+Q_{S}\rho-2Q_{S}\rho Q_{S}\big)\label{komME}
\eeq
This is a trace preserving master equation that describes the decoherence of the radical-pair's spin state brought about by the internal to the molecule recombination dynamics (the subscript in $d\rho_{\rm nr}$ denotes the "not reacted" molecules). This equation applies to a single quantum system. However, we have to model reactions, i.e. disappearance of radical-ion pairs and creation of neutral products. That was modelled by another equation evolving the existing number of radical-ion pairs,
\beq
dN=-Ndt(k_{S}\langle Q_{S}\rangle+k_{T}\langle Q_{T}\rangle)\label{dNdt}
\eeq
which is interpreted as follows: in the time interval between $t$ and $t+dt$ the probability of the radical-ion pair to be in e.g. the singlet state is $\langle Q_{S}\rangle=\tr\{\rho Q_{S}\}$, and hence the probability of recombination is $k_{S}dt\langle Q_{S}\rangle$. Since there exist $N(t)$ radical-ion pairs, $N(t)k_{S}dt\langle Q_{S}\rangle$ of them will disappear into singlet products, and similar considerations apply to the triplet channel, leading to the update rule for the number $N(t)$ shown in \eqref{dNdt}. As will be explained in the following, this description is part of the picture, and applies to states with maximal singlet-triplet coherence. The general case of incoherent mixtures is more complicated. In 2010 Jones \& Hore \cite{JH} introduced a different master equation derived from a rather ad-hoc application of the theory of generalized quantum measurements and positive maps. The equation reads
\begin{align}
d\rho/dt=-i[{\cal H},\rho]&-k_{S}\big(\rho Q_{S}+Q_{S}\rho-Q_{S}\rho Q_{S}\big)\label{JH}\\\nonumber &-
k_{T}\big(\rho Q_{T}+Q_{T}\rho-Q_{T}\rho Q_{T}\big)
\end{align}
\subsection{Conceptual problems of the Jones-Hore Theory}
The conceptual problem with the Jones-Hore theory is the erroneous application of single-quantum-system measurement dynamics to a many-particle ensemble of {\it independently} reacting molecules. In particular, the philosophy of the Jones-Hore derivation is that in a box with a macroscopic number of radical-ion pairs, the detection of the particular reaction products during the time interval $dt$ affects the quantum state of the remaining molecules. Specifically, these authors argue like this: if $k_{S}dt$ molecules react to produce singlet products, this will have the effect of {\it projecting} the rest of the system onto "a more triple" state, i.e. they density matrix acquires a term $k_{S}dtQ_{T}\rho Q_{T}$. This is an unphysical statement, for the following reasons: first, the molecules in the box are completely independent and non-interacting systems. The fact that molecule $i$ reacted {\it cannot} in any way affect the quantum state of molecule $j$. Second, in the quantum measurement of {\it single} quantum systems, one usually considers the quantum system and the measurement apparatus (another quantum system) coupled to it. Due to this coupling, the two become entangled, and of course any information extracted from the apparatus back-reacts on the system. In the case of radical-ion-pair reactions, we do not perform a global measurement on all molecules with some kind of apparatus. The measurement is internal in each molecule, as has already been explained \cite{kominis_PRE}. When a number of radical-ion pairs have reacted, the reaction products and the number thereof represent {\it purely classical information}. Whether or not we acquire this classical information is completely irrelevant and cannot have any back-reaction whatsoever on any quantum system. Finally, the authors have not derived their master equation (obviously because it is not derivable) using the usual recipe of considering the specific Hamiltonian interactions between system and environment and then tracing out the environment. Their ad-hoc postulated master equation following from a hand-waving use of the operator-product formalism completely lacks physical foundations. Moreover, as will be shown in Section V, the Jones-Hore theory is plagued by a severe inconsistency: if one applies it to a single-molecule experiment, the average of all possible realizations does not reproduce the predictions of the Jones-Hore master equation. The traditional approach faces the same problem, and both theories also violate conservation of energy, as will be shown in Section VI.

Most recently another three contributions appeared in the literature. Ivanov {\it et. al.} claimed to have "consistently" derived the traditional master equation, and Purtov has solved exactly a particular example and shown that the solution is what would follow from the traditional master equation. What both works have in common is the starting point of their derivation, which represents a deep physical misconception.
\subsection{The derivations of Ivanov {\it et. al.} and Purtov}
\begin{figure}
\includegraphics[width=8.5 cm]{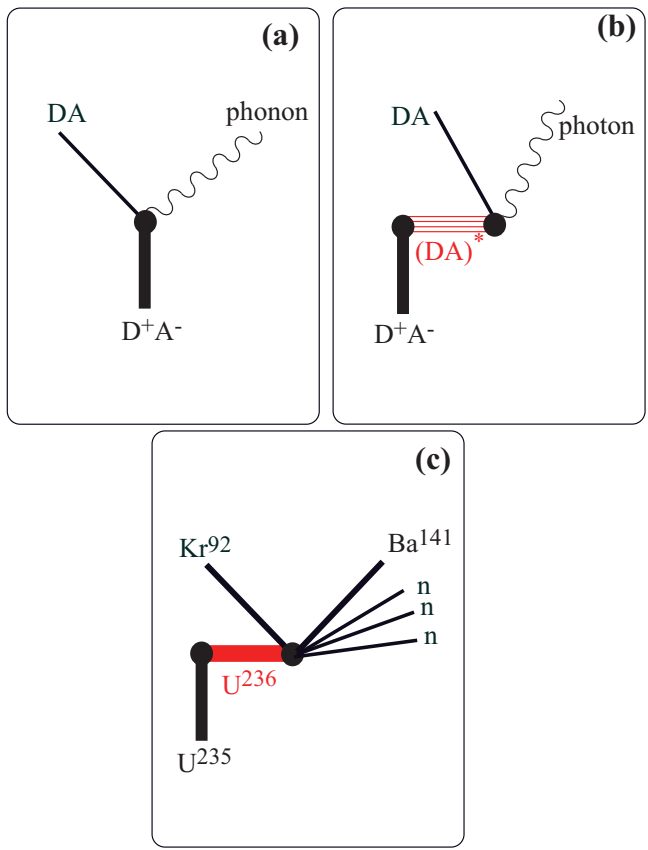}
\caption{Radical-ion-pair reactions pictured with Feynman diagrams. (a)The reaction as modelled by Ivanov {\it et. al.} and Purtov (b) The reactions as modelled by Kominis (c) Analogy with nuclear fission. It is the dense manifold of intermediate states that is ignored in the "effective" theory of Ivanov {\it et. al.} and Purtov. Similarly, it is the intermediate state of $^{236}$U that prohibits the formation of any quantum coherence between the parent nucleus and the fission product.}
 \label{fd}
\end{figure}

The starting point of both derivations is the assertion that the radical-ion pair quantum state, call it $|D^{+}A^{-}\rangle$ and the neutral product state, call it $|DA\rangle$, are eigenstates of one and the same Hamiltonian $H_{0}$, and transitions are induced between them by some perturbation $V$. That proposition, namely that reactants can be in a quantum superposition with the reaction products of an exothermic irreversible chemical reaction is utterly wrong. One can think of a similar situation, the nuclear decay of uranium into barium and krypton, and attempt to write for the total state $|\psi\rangle=c_{{\rm U}}|{\rm U}\rangle+c_{\rm Ba}|{\rm Ba}\rangle+c_{\rm Kr}|{\rm Kr}\rangle$, and continuing along these lines one could arrive at states of the form $|{\rm nuclear~bomb~exploded}\rangle+|{\rm nuclear~bomb~not~exploded}\rangle$. Or, to give an example familiar to chemists, one could argue along the lines of Ivanov {\it et. al.} that molecular oxygen, molecular hydrogen and water are eigenstates of one and the same Hamiltonian, and a perturbing potential causes transitions between them. If that would be the case, Nature would be dramatically different and we would not be reading this text now. Thus, the conceptual problem with these two derivations is that the authors build up a model that does not correspond to a physical entity. Hence their result is of no practical interest. Essentially, the authors build a so-called "effective" theory, ignoring a dense manifold of intermediate states and the {\it instant} decoherence caused by the {\it decay of these states}. In Feynman diagram terminology, these authors look at the radical-ion-pair reaction as a single vertex joining $|D^{+}A^{-}\rangle$ with $|DA\rangle$ and a phonon, as shown in Fig.\ref{fd}a.
The actual radical-ion-pair reaction, however, that takes place in nature, is described by the process shown in Fig.\ref{fd}b, where the radical-ion pair is converted into a vibrational excited state of the neutral molecule, ({\rm DA})$^{*}$, which then emits a photon and we then reach the ground state DA, which is the reaction product. The photon instantly "expels" to the environment the information on it's precursor state, and hence any coherence between the reactants and the reactions products {\it can not be physically sustained} and hence it is pointless and erroneous to consider it theoretically. The analogy with nuclear fission is shown in Fig.\ref{fd}c. In other words, to pertain that there exists a quantum state of the form $|\psi\rangle=\alpha|{\rm D}^{+}{\rm A}^{-}\rangle+\beta |{\rm DA}\rangle$ is a completely distorted view of physical reality. Philosophically, this viewpoint treats the whole universe as a single Hilbert space and induces coherent transitions between disjoint parts of this space. If nature indeed behaved like this, Schr{\" o}dinger cats would abound around us, in fact we ourselves would not exist in our current form to discuss this. We remind the reader that in the cold atom community various groups have created coherent oscillations between molecules and free atoms (see for example \cite{heinzen}), but at nano-Kelvin temperatures needed for Bose-Einstein condensation.  We are far (if ever) from doing the same in electron transfer or, for that matter, nuclear reactions.
\subsection{Shushin's Argument}
Shushin \cite{shushin} has argued that the Bloch-Redfield relaxation theory already explains radical-ion pair reaction dynamics and no new concept is required. The Bloch-Redfield theory is a second-order perturbation theory, and indeed similar terms like
$Q_{S}\rho$ or $Q_{S}\rho Q_{S}$ appear there as well. However, the theory describes decoherence induced by a noisy environment.
This has nothing to do with the charge recombination process. This is a fundamental decoherence process internal in the molecules. We cannot suppress this decoherence, no matter how good (i.e. noise free) experiments we perform. Shushin points out the main aspect of our previous work, namely the spin decoherence of not-yet-reacted radical-ion pairs, but regards it as an ad hoc interpretation. We claim that this interpretation directly follows from the formalism developed in \cite{kominis_PRE}, and its physical origin will be further elaborated upon in the following Section.
\section{Towards the Complete Theory}
In \cite{kom_short} we have outlined the derivation of the complete quantum dynamics description of radical-ion-pair reactions, while elucidating an analogy between this system and the double slit familiar from quantum optics. We will here elaborate on the complete theory in more detail. One part of the complete theory is the consistent treatment of the reaction. The other is the treatment of the not-yet-reacted radical-ion pairs, which undergo spin decoherence.
\subsection{Decoherence of Nonreacted Molecules}
\begin{figure}
\includegraphics[width=8.5 cm]{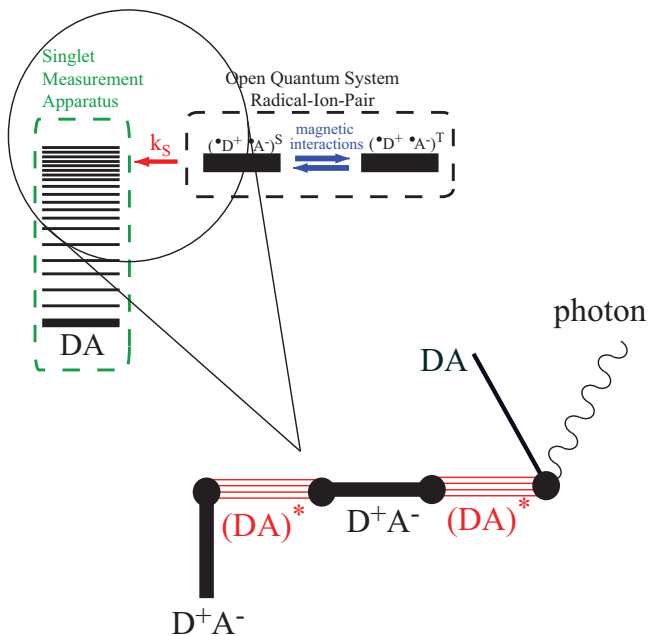}
\caption{A virtual process contributing to the decoherence of the radical-ion-pair spin density matrix. This decoherence is described by the density matrix equation derived by Kominis \cite{kominis_PRE}. In this virtual process, driven by the interaction Hamiltonian in the second order perturbation theory, the radical-ion pair couples to the excited vibrational state, which, instead of decaying to the ground state, couples back to the radical-ion pair. These processes are unobservable and the source of spin decoherence of nonreacted molecules.}
\label{fd2}
\end{figure}

What the master equation \eqref{komME} says is that the internal measurement going on in each radical-ion pair is responsible for spin decoherence. This internal measurement measures the observable $Q_S$ with a total measurement rate $k_{S}+k_{T}$, as explained in \cite{kominis_PRE}. A process contributing to loss of spin coherence can be nicely visualized by a Feynman diagram like the one shown Fig.\ref{fd2}. That is, a virtual tunneling from the radical-ion pair state to the singlet reservoir state {\it and back} has the effect to dephase the coherent-singlet triplet oscillation. Similar is the effect of the triplet reservoir. Such virtual processes are unobservable. This is the physical interpretation of tracing out the reservoir degrees of freedom, while assuming a still-existing radical-ion pair, as explained in \cite{kominis_PRE}.
\subsection{Maximum Coherence Extreme: Problem with the Traditional and Jones-Hore Theory}
In all subsequent discussions we neglect nuclear spins and focus on a two-level radical-ion pair, the Hilbert space of which is spanned by two states,
the singlet $\|S\rangle$ and the triplet $|T\rangle$ (we should say the triplet $|T_{0}\rangle$, the zero projection state of the triplet manifold spanned by $|T_{\pm}\rangle$ and $|T_{0}\rangle$, but for notational simplicity we write $|T\rangle$).

Suppose we have a macroscopic number $N$ of radical-ion pairs. Each is described by a density matrix $\rho_{i}$, hence the density matrix of the ensemble is $\rho=\sum_{i=1}^{N}{\rho_{i}}$. Suppose that particle $j$ recombines into the singlet channel. What is the effect of this recombination on $R$? Clearly, now $R=\sum_{i\neq j}{\rho_{i}}$. That is, we have to remove one particle from the ensemble of particles, and from the total quantum state $R$ we have to remove $\rho_{j}$. What was the state of the particle to be removed? Was it the singlet state $|S\rangle\langle S|$ because we found a singlet product? Not necessarily. To claim that is a gross misunderstanding of the state reduction. Consider for simplicity a single two level atom observed by a unit efficiency photo-detector. If the detector clicks, i.e. a photon is observed, we know for sure that the atom is ({\it now, just after the click}) in the ground state $|g\rangle$. We know nothing as to what state the atom was in before observing the photon. That is, we cannot say that just before observing a photon the atom was for sure in the excited state $|e\rangle$. The atom could have been at any quantum superposition of the form $\alpha|g\rangle+\beta|e\rangle$. In other words, the observation of the photon {\it updates} our knowledge about the state of the atom, it does not inform us about the atomic state prior to the observation. Similarly, the observation of a singlet product molecule does not at all mean that just prior to the observation the radical-ion pair was in the singlet state. It might have been in any coherent superposition of singlet and triplet states, and the molecule decided to recombine through the singlet channel.

Suppose now that all $N$ radical-ion pairs are at time $t$ prepared in the same state $|\psi\rangle=\alpha|S\rangle+\beta|T\rangle$, hence the single-molecule density matrix is $\rho_{1}=|\alpha|^{2}|S\rangle\langle S|+|\beta|^{2}|T\rangle\langle T|+\alpha\beta^{*}|S\rangle\langle T|+\alpha^{*}\beta|T\rangle\langle S|$. The ensemble density matrix is $\rho=N\rho_{1}$. Clearly, in the time interval between $t$ and $t+dt$ we will detect $dn_{S}=k_{S}dt\tr\{\rho Q_{S}\}=k_{S}dtN|\alpha|^{2}$ singlet and $dn_{T}=k_{T}dt\tr\{\rho Q_{T}\}=k_{T}dtN|\beta|^{2}$ triplet neutral products. The change in $\rho$ will then be $d\rho=\rho_{1}dN+Nd\rho_{1}$. That is,
$\rho$ changes because radical-ion pairs react (first term) and because non-reacted radical-ion pairs are subject to a state change. From the work of Kominis \cite{kominis_PRE} it follows that this state change is singlet-triplet decoherence, and it is described by \eqref{komME}.
Let us now see what each of the theories has to say about $d\rho$.
\subsubsection{Kominis Theory}
From \eqref{dNdt} we find that $dN=-dn_{S}-dn_{T}$, whereas from \eqref{komME} we find that $d\rho_{1}=-(k_{S}+k_{T})dt\rho_{\rm coh}/2$, where $\rho_{\rm coh}=N(\alpha\beta^{*}|S\rangle\langle T|+\alpha^{*}\beta|T\rangle\langle S|)$ is the coherence of $\rho$. Hence in this theory we find
\beq
d\rho=-(dn_{S}+dn_{T})\rho_{1}-{{(k_{S}+k_{T})dt}\over 2}\rho_{\rm coh}
\eeq
\subsubsection{Traditional Theory}
From \eqref{phen} it is straightforward to find that
\begin{align}
d\rho&=-(dn_{S}+dn_{T})\rho_{1}+({{dn_{S}+dn_{T}}\over N}-{{(k_{S}+k_{T})dt}\over 2})\rho\nonumber\\
&-{1\over 2}\Big((dn_{S}-dn_{S}^{'})|S\rangle\langle S|+(dn_{T}-dn_{T}^{'})|T\rangle\langle T|\Big)
\end{align}
where $dn_{S}^{'}=k_{T}|\alpha|^{2}$ and $dn_{T}^{'}=k_{S}|\beta|^{2}$.
\subsubsection{Jones-Hore Theory}
From \eqref{JH} we find that
\begin{align}
d\rho&=-(dn_{S}+dn_{T})\rho_{1}+\Big({{dn_{S}+dn_{T}}\over N}-(k_{S}+k_{T})dt\Big)\rho\nonumber\\
&+dn_{S}^{'}|S\rangle\langle S|+dn_{T}^{'}|T\rangle\langle T|
\end{align}
\subsubsection{Physical Problem of the Traditional and the Jones-Hore Theories}
The common problem of the traditional and the Jones-Hore theory is in the second term of their expressions for $d\rho$. Namely both theories update the quantum state of the nonreacted molecules based on the recombination events of those molecules that reacted. This unphysical state of affairs has been analyzed in \cite{kom_short} in analogy with the photon double slit experiment. We briefly reiterate the main point: the fact that in the time interval $dt$ we have $dn_{S}$ singlet and $dn_{T}$ triplet recombinations affects the density matrix of the non-reacted molecules, as if there is a weird communication between radical-ion pairs in the ensemble, is unphysical, as unphysical is to assert that the detection of the number of photons on the observation screen in Young's double slit experiment affects the quantum state of the other photons flying through the slits. In an ensemble of 1000 say radical-ion pairs being in a coherent singlet-triplet superposition, the fact that 10 of them recombined into the singlet channel in the time interval between $t$ and $t+dt$ cannot have any influence on the spin state of the other 990 whatsoever (unless they are entangled, which is a situation far from possible at the current experiments).
In contrast, according to the theory of Kominis, the nonreacted molecules suffer loss of their spin coherence as time goes on, which is {\it a single molecule effect acting internally and independently in each molecule}.
\subsection{Maximum Incoherence Extreme: Problem with the Kominis Theory}
A maximally coherent state is one extreme possibility. The other extreme is the maximally incoherent state. We will study this
case now and then move on to the more realistic general case. Suppose that at some instant in time we know that the
ensemble density matrix is an equal mixture between singlet and triplet, i.e. suppose that
\beq
\rho_{t}={N\over 2}|S\rangle\langle S|+{N\over 2}|T\rangle\langle T|=N\rho_{1}
\eeq
where $\rho_{1}=(|S\rangle\langle S|+|T\rangle\langle T|)/2$. That is, according to $\rho_t$ there are $N/2$ radical-ion pairs in the singlet subspace and another $N/2$ in the triplet. We again consider the absence of magnetic interactions, so there is no singlet-triplet mixing, and we consider just one recombination channel, for example we take $k_{S}=0$. What we naturally expect to find is that half of the radical-pairs will disappear into triplet products, leaving us at the end with a density matrix $\rho_{\infty}={N\over 2}|S\rangle\langle S|$ describing the other half which have remained in the non-reactive singlet state.
\subsubsection{Kominis Theory}
As before, $d\rho=\rho_{1} dN+Nd\rho_{1}$. Since $\rho_{1}=(Q_{S}+Q_{T})/2$ it easily follows that $d\rho_{\rm nr}=0$ i.e. there is no more coherence to be lost from an incoherent mixture, hence the master equation \eqref{komME} for nonreacted molecules does not induce any change in $\rho$.
Since furthermore $\langle Q_{S}\rangle=\langle Q_{T}\rangle=1/2$, the equation determining the number of radical-ion pairs \eqref{dNdt} gives
$dN=-k_{T}dtN/2$, hence we get $d\rho/dt=-k_{T}\rho/2$ i.e. $\rho_{\infty}=0$ (meaning all radical-ion pairs disappeared into products), which obviously is not the case. So it is clear that the machinery of removing particles according to \eqref{dNdt} does not work in this case \cite{hore}. The reason is that, writing $\rho_{t}=N\rho$, upon measuring $dn_{T}$ triplet recombination products we should update $\rho_t$ by the rule $\rho_{t+dt}=\rho_{t}+d\rho$, where $d\rho=-dn_{T}Q_{T}$, whereas Kominis' theory updates $\rho_t$ by $d\rho=-dn_{T}\rho$. In other words, in this extreme incoherent case {\it we know for sure} that the observation within the time interval $dt$ of $dn_{T}$ triplet products implies that these products came {\it from triplet state radical-pairs}, so our knowledge about the ensemble state is updated to a state having "less triplet character", i.e. we have to remove from $\rho$ the projection $dn_{T}Q_{T}\rho Q_{T}$.
\subsubsection{Traditional and Jones-Hore Theories}
It is straightforward to show that both the phenomenological as well as the Jones-Hore theories gives
\beq
d\rho=-k_{T}dt{{NQ_{T}}\over 2}
\eeq
leading to $\rho_{\infty}=NQ_{S}/2$, as it should be. Thus in this, maximum incoherence extreme, both of these theories work fine.
\section{General Case}
In \cite{kom_short} we developed the formalism addressing the general case, which is this: suppose that at time $t$ the radical-ion-pair ensemble is described by the density matrix $\rho_{t}$, which can be any mixture. In the following time interval $dt$ we will surely find $dn_{S}=k_{S}dt\tr\{\rho Q_{S}\}$ singlet and $dn_{T}=k_{T}dt\tr\{\rho Q_{T}\}$ triplet neutral products. How can we consistent with all information at hand update $\rho_{t}$ into $\rho_{t+dt}=\rho_{t}+d\rho$? For completeness, we reiterate the answer to this question. If $\rho_{t}$ were an incoherent mixture of singlet and triplet radical-ion pairs, then we could independently remove "singlet" and "triplet" character from $\rho$ according to $d\rho_{\rm incoh}=-k_{S}dtQ_{S}\rho Q_{S}-k_{T}dtQ_{T}\rho Q_{T}$. In the other extreme, if $\rho_{t}$ was a state with maximal singlet-triplet coherence, then
we would have to remove the complete information on the state of the reacted molecules, which is what we did in our previous work \cite{kominis_PRE}: $d\rho_{\rm coh}=-(dn_{S}+dn_{T})\rho_{t}/\tr\{\rho_{t}\}$, where $\rho_{t}/\tr\{\rho_{t}\}$ is the single-molecule density matrix. In the general case we have at our disposal a measure of singlet-triplet coherence of $\rho_{t}$. This measure was defined in \cite{kom_short} as
\beq
p_{\rm coh}={{\tr\{\rho_{ST}\rho_{TS}\}}\over {\tr\{\rho_{SS}\}\tr\{\rho_{TT}\}}}
\eeq
where $\rho_{SS}=Q_{S}\rho Q_{S}$, $\rho_{TT}=Q_{T}\rho Q_{T}$, $\rho_{ST}=Q_{S}\rho Q_{T}$ and $\rho_{TS}=Q_{T}\rho Q_{S}$.
The incoherent part of $\rho$ is $\rho_{\rm incoh}=\rho_{SS}+\rho_{TT}$, while the coherence is $\rho_{\rm coh}=\rho_{ST}+\rho_{TS}$. More on $p_{\rm coh}$ in the following. Finally, we also have to add the change in the density matrix of the not-yet-reacted radical-ion pairs, $d\rho_{\rm nr}$, given by \eqref{komME}. Putting everything together we get
\begin{widetext}
\beq
{{d\rho}\over {dt}}={{d\rho_{\rm nr}}\over {dt}}-(1-p_{\rm coh})(k_{S}Q_{S}\rho Q_{S}+k_{T}Q_{T}\rho Q_{T})
-p_{\rm coh}\big(k_{S}\tr\{Q_{S}\rho\}+k_{T}\tr\{Q_{T}\rho\}\big){\rho\over {\tr\{\rho\}}}\label{KOM}
\eeq
\end{widetext}
\subsection{Reaction of Spin Coherent Radical-Ion Pairs}
\begin{figure}
\includegraphics[width=6.0 cm]{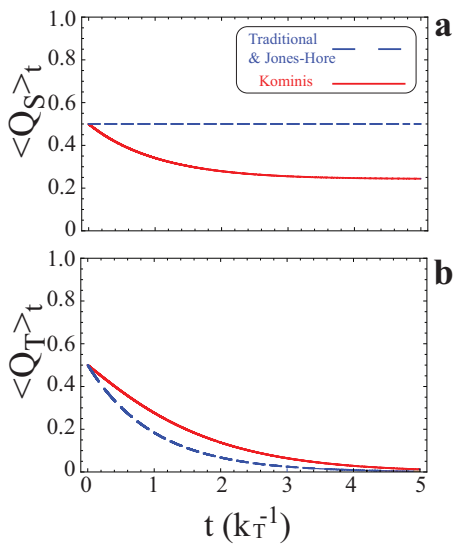}
\caption{Radical-ion-pair recombination dynamics in the absence of singlet-triplet mixing and $k_{S}=0$, starting from the initial state $(|S\rangle+|T\rangle)/\sqrt{2}$. (a) Time evolution of $\tr\{Q_{S}\rho\}$ and (b) of $\tr\{Q_{T}\rho\}$ as predicted by \eqref{KOM} (red solid line) and the traditional as well as the Jones-Hore theory (blue dashed line).}
\label{zeroHam}
\end{figure}
To illustrate the fundamental differences between the various theories, we used \cite{kom_short} a very simple scenario of an ensemble of radical-ion pairs all starting out in the coherent state $(|S\rangle+|T\rangle)/\sqrt{2}$, with no singlet-triplet mixing (i.e. we zero the magnetic Hamiltonian) and just one recombination channel (e.g. we set $k_{S}=0$).  It is straightforward to find that both the traditional as well as the Jones-Hore theory predict that half of the radical pairs will react (of course producing triplet products) while the other half will remain locked in the non-reacting singlet state (Fig.\ref{zeroHam}).
In glaring contrast, the newly developed theory predicts that 75\% of the radical pairs will react, and 25\% will remain in the singlet state forever. At first look this sounds completely counter-intuitive and almost unacceptable, according to the layman understanding of quantum mechanics. Since the initial state is a coherent superposition of singlet and triplet with equal amplitudes and the singlet is non-reacting, how on earth do 75\% of the molecules react? The resolution of this apparent paradox is this: the traditional theory is based on a derivation like the one presented by Ivanov {\it et. al.} \cite{ivanov}. This theoretical construct allows the following evolution: Starting with a radical-ion-pair state ${1\over \sqrt{2}}(|S\rangle+|T\rangle)$, and because the singlet is non-reactive and the triplet reactive, this state evolves to
${1\over \sqrt{2}}(|{\rm singlet~radical~pair}\rangle+|{\rm triplet~product}\rangle)$ and of course that results in 50\% triplet products!! But such an evolution is impossible. As we explained previously, the reaction {\it does not} create coherent superpositions between reactants and reaction products. Asking an ill-defined question in quantum physics can lead to paradoxical results. Simple projective-measurement quantum mechanics says that when we measure a physical observable $A$ in a state which is a coherent superposition of the observable eigenstates, say $|\psi\rangle=c_{1}|a_{1}\rangle+c_{2}|a_{2}\rangle$ we obtain $a_{1}$ with probability $|c_{1}|^{2}$ and so on. So a question that can be asked regarding the initial state ${1\over \sqrt{2}}(|S\rangle+|T\rangle)$ is this: what is the probability in a projective measurement of $Q_{S}$ to find zero or one? Of course both probabilities are 1/2. But this does not tell us anything about the reaction products. This is so because the radical-ion pair might be in this coherent state at time $t$ and there is a finite probability of reaction within the following time interval $dt$. But the radical-ion pair  might choose not to react, and its spin state will then evolve according to \eqref{komME}, hence the final product yield is not that straightforward to predict by using the naive projective measurement argument. To reiterate, there is no observable having a radical-ion-pair state and a neutral product state as eigenstates. This is the source of the confusion and the apparently paradoxical result we mentioned before.
\subsection{Single Molecule Scenarios}
We will here analyze single molecule experiments to further illuminate the deficiencies of the traditional and Jones-Hore theories. Consider a single radical-ion pair at time $t$ in the coherent superposition state ${1\over \sqrt{2}}(|S\rangle+|T\rangle)$. Again, we eliminate singlet-triplet mixing and take $k_{S}=0$. Clearly, $\langle Q_{S}\rangle_{t}=\langle Q_{T}\rangle_{t}=1/2$. Within the next time interval $dt$, there is a finite probability $dp_{T}=k_{T}dt\langle Q_{T}\rangle_{t}=k_{T}dt/2$ of recombination in the triplet channel. In a particular realization of this experiment that probability will materialize, so at time $t+dt$ we have no radical-ion pairs, hence $\langle Q_{S}\rangle_{t+dt}=\langle Q_{T}\rangle_{t+dt}=0$. However, both the traditional as well as the Jones-Hore theory
predict $\langle Q_{S}\rangle_{t}=1/2$ for all times. How are these two contradictory facts reconciled? Well, we have to consider a different realization of this single molecule experiment, in which the radical-ion pair does not react within the time interval $dt$ after its preparation in the coherent state at time $t$. How does the state of an nonreacted molecule evolve according to the Jones-Hore theory? In their derivation of the Jones-Hore master equation \cite{JH}, the authors argue: during the time interval $dt$ we will have $k_{S}dt$ radical pairs recombine through the singlet channel, making the rest more "triplet", i.e. adding to $\rho$ the term $k_{S}dtQ_{T}\rho Q_{T}$. Similarly we will have another $k_{T}dt$ radical pairs recombine through the triplet channel, adding to $\rho$ the term $k_{T}dtQ_{S}\rho Q_{S}$. And finally, we add to $\rho$ the term $(1-k_{S}dt-k_{T}dt)\rho$ because a fraction $1-k_{S}dt-k_{T}dt$ of radical-ion pairs did not recombine at all. Thus the update rule for the Jones-Hore master equation is $\rho_{t+dt}=(1-k_{S}dt-k_{T}dt)\rho+k_{S}dtQ_{T}\rho Q_{T}+k_{T}dtQ_{S}\rho Q_{S}$, which leads to \eqref{JH}. So according to this theory the density matrix of a non-reacting radical-ion pair will change by $d\rho=0$. In other words, the state of a non-reacting radical-ion pair remains ${1\over \sqrt{2}}(|S\rangle+|T\rangle)$ until the time it reacts, i.e. $d\langle Q_{S}\rangle=0$. But in the previous realization we had $d\langle Q_{S}\rangle=-1/2$, so to keep $d\langle Q_{S}\rangle=0$, we must have realizations with d$\langle Q_{S}\rangle>0$, so that the average satisfies the prediction of \eqref{JH} shown in Fig.\ref{zeroHam}a. So it is clear that the Jones-Hore theory is not self-consistent, i.e. it cannot account for single molecule scenarios in such a way that the average of them reproduces the predictions of the Jones-Hore master equation \eqref{JH}. To resolve this problem and keep $d\langle Q_{S}\rangle=0$ the Jones-Hore theory performs a rather unphysical act, which {\it necessitates} the existence of more than one radical-ion pairs: in an ensemble of molecules $N$, the fact that a number of them reacts through the triplet channel {\it projects the rest to a state with more singlet character}. Although we are repeating ourselves here, it is a crucial point and we will reiterate: the fact that in our box containing $10^{23}$ radical-ion pairs a number of them (say $10^6$) recombined within the time interval $dt$ to create triplet products, this very fact {\it projects} the rest $10^{23}-10^{6}$ molecules in a state that is more singlet than the state they were at time $t$. We claim that this is an impossible act. To reiterate, the Jones-Hore theory fails to account for single-molecule realizations in a self-consistent way, and manages to keep $d\langle Q_{S}\rangle=0$ by some weird action at a distance between independent molecules in an ensemble. The traditional theory does not even in principle consider what happens to a non-reacting molecule. Such a concept does not exist in the traditional theory, therefore the above considerations apply to the traditional theory as well.

\begin{figure}
\includegraphics[width=8.0 cm]{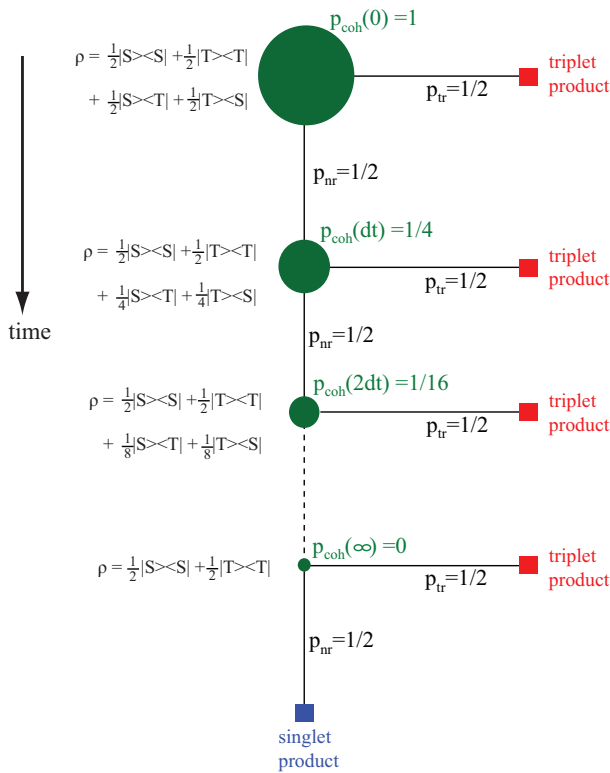}
\caption{Single radical-ion-pair state evolution with no singlet-triplet mixing, a single recombination channel ($k_{S}=0$) and a coherent initial state. The probability for triplet recombination is $p_{\rm tr}=1/2$ and the probability for no recombination is $p_{\rm nr}=1/2$.}
\label{diagram}
\end{figure}

In contrast, the picture emerging from our description is the following: Take $k_{T}dt=1$, so that starting with the coherent state ${1\over \sqrt{2}}(|S\rangle+|T\rangle)$ the probability for triplet recombination is $k_{T}dt/2=1/2$. We will trace the state evolution at time steps $dt$, as shown in Fig.\ref{diagram}. In a particular realization of this single molecule experiment this recombination takes place, and the reaction terminates.
In another realization the radical-ion pair will not react in this time interval $dt$ after its preparation in the state $|\psi\rangle$, so according to \eqref{komME} spin coherence will be lost. At time $t=0$ we had $p_{\rm coh}=1$, but if the radical-ion pair does not react, it will lose half of its coherence (easily found from \eqref{komME}) and at time $t=dt$ the coherence measure $p_{\rm coh}$ will be reduced to $p_{\rm coh}=1/4$. At this point there is again the possibility of triplet recombination, the probability of which is still 1/2. Again, if this recombination does not take place, coherence will be still reduced, and at time $t=2dt$ we will have $p_{\rm coh}=1/16$. All these are possible realizations, as is the the realization that we have no reaction until the coherence has been reduced to exactly zero, where we have an incoherent single-triplet mixture. At this point the radical-ion pair will either remain in the non-reacting singlet state (probability 1/2) or recombine in the triplet channel (probability 1/2). So the longest the reaction can proceed is until $d\rho_{\rm nr}/dt=0$. The question is, however, when is the coherence really zero? In other words, how many realizations do we have to average in order to estimate the triplet yield? Again $p_{\rm coh}$ is the measure of when the reaction is really over. So the triplet yield $Y_T$ is not $p_{\rm tr}+p_{\rm nr}p_{\rm tr}+p_{\rm nr}^{2}p_{\rm tr}+...={1\over 2}+{1\over 2^2}+...+{1\over 2^n}$, but each term in the series is suppressed by $p_{\rm coh}$, i.e.
\beq
Y_{T}=p_{\rm tr}+p_{\rm coh}(0)p_{\rm nr}p_{\rm tr}+p_{\rm coh}(dt)p_{\rm nr}^{2}p_{\rm tr}+...
\eeq
This is how the result shown in Fig.\ref{zeroHam}a is produced.
\subsection{More on $p_{\rm coh}$}
The meaning of the definition of $p_{\rm coh}$ is rather obvious if we consider a mixture of $N_{c}$ radical-ion pairs in the coherent state $\alpha|S\rangle+\beta|T\rangle$, $N_{S}$ radical-ion pairs in the singlet state $|S\rangle$ and $N_{T}$ radical-ion pairs in the triplet state $|T\rangle$. The density matrix of such a mixture is $\rho=(N_{S}+N_{c}|\alpha|^{2})|S\rangle\langle S|+(N_{T}+N_{c}|\beta|^{2})|T\rangle\langle T|+N_{c}\alpha\beta^{*}|S\rangle\langle T|+N_{c}\alpha^{*}\beta|T\rangle\langle S|$.
In this case $\tr\{\rho_{ST}\rho_{TS}\}=N_{c}^{2}|\alpha|^{2}|\beta|^{2}$, $\tr\{\rho_{SS}\}=N_{S}+N_{c}|\alpha|^{2}$ and $\tr\{\rho_{TT}\}=N_{T}+N_{c}|\beta|^{2}$. As intuitively expected, if $N_{c}=0$ we have $p_{\rm coh}=0$, and if $N_{S}=N_{T}=0$ we get $p_{\rm coh}=1$. Now let us take into account the nuclear spin degrees of freedom. For simplicity we consider just one nuclear spin with spin 1/2 and we denote by $|\pm\rangle$ the nuclear spin basis states. If the radical-ion-pair spin state is for example ${1\over\sqrt{2}}(|S\rangle+|T\rangle)|+\rangle$, then again $p_{\rm coh}=1$. If, on the other hand the nuclear spin is in an incoherent mixture of $|+\rangle$ and $|-\rangle$, then the spin state of the radical-ion pair will be described by the density matrix $\rho={1\over 4}(|S\rangle\langle S|+|T\rangle\langle T|+|S\rangle\langle T|+|T\rangle\langle S|)(|+\rangle\langle +|+|-\rangle\langle -|)$ and we find that $p_{\rm coh}=1/2$. In this case the electronic spin is in a maximally coherent state for which we expect $p_{\rm coh}=1$ but the nuclear spin is in mixed state, and therefore $p_{\rm coh}=1/2$ underestimates the electron spin coherence. This is because $p_{\rm coh}$ depends on $\tr\{\rho_{ST}\rho_{TS}\}$, which is also a measure of the mixed character of the spin state. However, it is just electron spin coherence that we care about regarding the recombination process. We therefore have to normalize $\tr\{\rho_{ST}\rho_{TS}\}$ by $\tr\{\rho^{2}\}$, and accordingly we normalize $\tr\{\rho_{SS}\}$ and $\tr\{\rho_{TT}\}$ by $\tr\{\rho\}$. Now $p_{\rm coh}=1$ for the aforementioned mixed nuclear spin state, as it should be.
\section{Energy Conservation in Radical-Ion-Pair Reactions}
\begin{figure}
\includegraphics[width=7 cm]{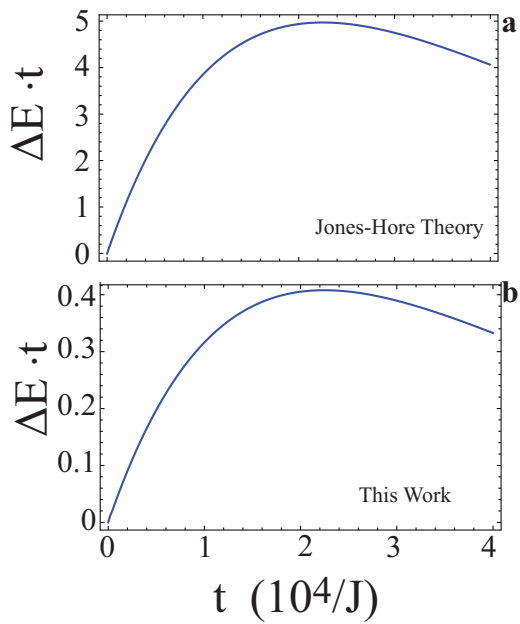}
\caption{Evolution of $\Delta E(t)$ predicted by (a) the Jones-Hore theory and (b) this work, i.e. equation \eqref{KOM}. We consider a radical-ion pair starting in the state $|++-\rangle$, i.e. the two electrons are in sth spin up state and the single nuclear spin in the spin down state. We take $k_{S}=0$, $k_{T}=1.0$, the magnetic field is zero, the hyperfine coupling with the single nucleus is $a=1.0$ and the exchange coupling $J=20.0$.}
\label{energy}
\end{figure}

In radical-ion-pair reactions, singlet-triplet mixing conserves energy, that's why the magnetic Hamiltonian mixes the singlet with the degenerate $T_{0}$ state of the triplet manifold. It is well known that the exchange interaction, of the form
$-J\mathbf{s}_{1}\cdot\mathbf{s}_{2}$ splits the singlet and triplet manifolds by $J$ and hence suppresses singlet-triplet mixing.
We study this effect and compare how the three theories deal with energy conservation. We consider a radical-ion pair with a single spin-1/2 nucleus and take as an initial state $|++-\rangle$, i.e. the two electrons (first two entries) are in the $T_{+}$ state and the nuclear spin $\mathbf{I}$ is in the $|-\rangle$ state. We consider an isotropic hyperfine coupling $a\mathbf{s}_{1}\cdot\mathbf{I}$ of one electron with the single nucleus and we also consider the exchange coupling between the electrons, hence the magnetic Hamiltonian is ${\cal H}=-J\mathbf{s}_{1}\cdot\mathbf{s}_{2}+a\mathbf{s}_{1}\cdot\mathbf{I}$. Since during the reaction we lose radical-ion pairs into reaction products, we consider the quantity
\beq
\Delta E(t)\equiv \tr\{\rho_{t}{\cal H}\}-\tr\{\rho_{t=0}{\cal H}\}\tr\{\rho\}_{t}
\eeq
That is, we consider the expectation value of the system's energy, $\tr\{\rho{\cal H}\}_{t}$, which obviously tends to zero due to the reaction removing molecules, minus the energy the system would have at time $t$ if the energy per molecule remained the same. Stated otherwise, $\Delta E(t)$ is the change in the energy per radical-ion-pair times the number of radical-ion pairs. Since in our considerations regarding the fundamental dynamics the radical-ion pair {\it does not} exchange energy with the environment, we expect $\Delta E(t)=0$, at least within $1/t$, i.e. energy should be conserved, or at least, if it is not, this non-conservation should be within the limits allowed by Heisenberg's uncertainty. The difference $\Delta E(t)$ is given units of the exchange coupling $J$, whereas time has units of $1/J$. The product $\Delta E(t) t$ should thus not exceed unity. However, this is the case with the traditional and Jones-Hore theories, in contrast with the result following from the general theory developed here and in \cite{kom_short}. This is shown in Fig.\ref{energy}.
It is seen that the Jones-Hore theory (the same result holds for the traditional theory) violates energy conservation by 5$\sigma$, whereas our description keeps the system's energy change within the physical limits.
\subsection{Analogy with Quantum Dots}
A similar issue comes up in the system of two coupled quantum dots interrogated by a point contact, succinctly analyzed by Gurvitz \cite{gurvitz}. The tunneling rate in the point contact, $T$, depends on whether the left or the right dot is occupied, while a coherent oscillation takes place between the two extreme states (a) left well occupied and (b) right well occupied. Using Bloch equations, Gurvitz describes the decoherence brought about by the measurement device of the double-well coherent oscillations. There are three rates that determine the problem: the coherent oscillation frequency $\Omega_{0}$, the energy mismatch $\epsilon$ of the ground state energies of the two wells and the decoherence rate $\Gamma$, derived from the point contact tunneling rates. When the decoherence rate $\Gamma=0$, the coherent oscillations are evident in the probability of occupation of the left well. For increasing $\Gamma$, the coherent oscillations are suppressed and the density matrix tends to the fully mixed state (in this case it is a two-dimensional density matrix and hence $\sigma_{aa}\rightarrow 1/2$ as $t\rightarrow\infty$). The first point of relevance to radical-ion pairs is that, as Gurvitz notes, this is a counter-intuitive manifestation of the quantum Zeno effect, namely that the effect of a strong continuous measurement is to delocalize the state of the measured system, as opposed to the more familiar localization produced by a series of projective measurements. The exact analog with radical-ion pairs is the decoherence of the nonreacted molecules. The evolution $d\rho_{\rm nr}/dt$ brings about exactly such a delocalization. The second point is that there is a substantial probability to occupy the right well (50\% as $t\rightarrow\infty$), even though {\it the energies of the two wells differ by} $\epsilon\neq 0$. That is, the electron tunneling from the left to the right well seems not to conserve energy. This apparent paradox is easily resolved, since the measurement apparatus, the quantum point contact, can readily exchange energy with the quantum system. Clearly so, because the interaction Hamiltonian coupling the left well with the point contact is essentially an electrostatic interaction, embodied in the
"Coulomb-blockade"-like term
${\cal H}_{\rm int}=\sim{\delta\Omega_{lr} c_{L}^{\dagger}c_{L}(a_{l}^{\dagger}a_{r}+{\rm h.c.})}$
where $c_{L}$ ($c_{L}^{\dagger}$), $a_{l}$ ($a_{l}^{\dagger}$) and $a_{r}$ ($a_{r}^{\dagger})$ are annihilation (creation) operators for the left well, the left reservoir and the right reservoir of the point contact barrier, respectively. In other words, the electron on the left well interacts with the point contact and can hence make a transition with $\Delta E\neq 0$. In radical-ion pairs, however, the coupling with the "environment" is the tunneling analyzed in \cite{kominis_PRE}, the result of which is a pure decoherence. This is so because there is no mechanism for the radical-ion pair to exchange energy with the vibrational reservoir, hence the spin state evolution must conserve energy.
\section{Conclusions}
We have analyzed in detail two extreme cases, that of maximal singlet triplet coherence, and that of maximal incoherence. The fundamental theories of radical-ion pair reactions that existed so far were applicable only at one of the two extremes. We have developed the general theory that spans the continuous range between these two extremes. Furthermore, we have elucidated the fundamental conceptual problems faced by the traditional master equation of spin chemistry as well as the more recent variant by Jones \& Hore. We have unraveled the inconsistencies of these theories by attempting to describe single-molecule measurement scenarios. Finally we have compared all theories in the front of energy conservation. In all cases the general theory we have developed consistently describes several gedanken-experiments. It is clear, as noted in \cite{kom_short}, that when the singlet-triplet coherence (quantified by $p_{\rm coh}$) is small, the traditional theory will provide a satisfying description of spin-selective radical-ion pair reactions. Such a case would be experiments dominated by spin relaxation. On the other hand, in experiments suppressing relaxation effects to the extent that $p_{\rm coh}$ is large, we expect significant deviations from the previous theoretical understanding.


\begin{thebibliography}{0}
\bibitem{hore}
The problem that the reaction kinetics as described by \eqref{dNdt} do not lead to sensible results in this example of an incoherent
mixture without singlet-triplet mixing and just one recombination channel was brought up by Prof. P. Hore.
\end{thebibliography}
\end{document}